\documentclass[reprint,aps,prb,floatfix]{revtex4-2}
\usepackage{bm}
\usepackage{amssymb}
\usepackage{amsmath}
\usepackage{graphicx}
\usepackage{rotating}
\usepackage{epsfig}
\usepackage{psfrag}
\usepackage{amsmath}
\usepackage{subfigure}
\usepackage{braket}
\usepackage{tikz}
\usepackage{stmaryrd}
\usepackage{wasysym}

\newcommand{\bea}{\begin{eqnarray}}
\newcommand{\eea}{\end{eqnarray}}
\newcommand{\bpm}{\begin{pmatrix}}
\newcommand{\epm}{\end{pmatrix}}

\usepackage[unicode=true,colorlinks=true,pdfpagemode=UseOutlines,pdfstartview=Fith]{hyperref}
\hypersetup{linkcolor=blue,citecolor=blue,urlcolor=blue}

\begin{document}
\title{Nematic heavy fermions and coexisting magnetic order in CeSiI}

\author{Aayush Vijayvargia$^1$, Onur Erten$^1$}
\affiliation{$^1$Department of Physics, Arizona State University, Tempe, AZ 85287, USA}

\begin{abstract}
Motivated by the recent discovery of magnetism and heavy quasiparticles in van der Waals material CeSiI, we develop an effective model that incorporates the conduction electrons residing at the silicene layer interacting with the local moments of the Ce ions. Ce sites are arranged on two layers of triangular lattices, above and below the silicene layer, and they are located at the center of the honeycomb lattice. This arrangement results in an effective extended Kondo interaction along with a predominant ferromagnetic RKKY interaction. Via mean-field theory of Abrikosov fermions, our analysis indicates that the ground state of the monolayer can exhibit a non-magnetic nematic heavy fermion phase that breaks C$_6$ rotational symmetry for small Heisenberg exchange and a magnetically ordered phase for large Heisenberg exchange. For intermediate values, a coexistence of magnetic order and a uniform heavy Fermi liquid is stabilized where they reside on separate Ce layers. We show that this phase can further be enhanced by an external electric field. Our results provide a natural mechanism for the coexistence of magnetic order and heavy fermions in CeSiI and highlight the possibility of unconventional non-magnetic heavy fermions with broken rotational symmetry.
\end{abstract}
\maketitle
\section{Introduction}
\begin{figure}[t]
    \centering
    \includegraphics[width=0.8\linewidth]{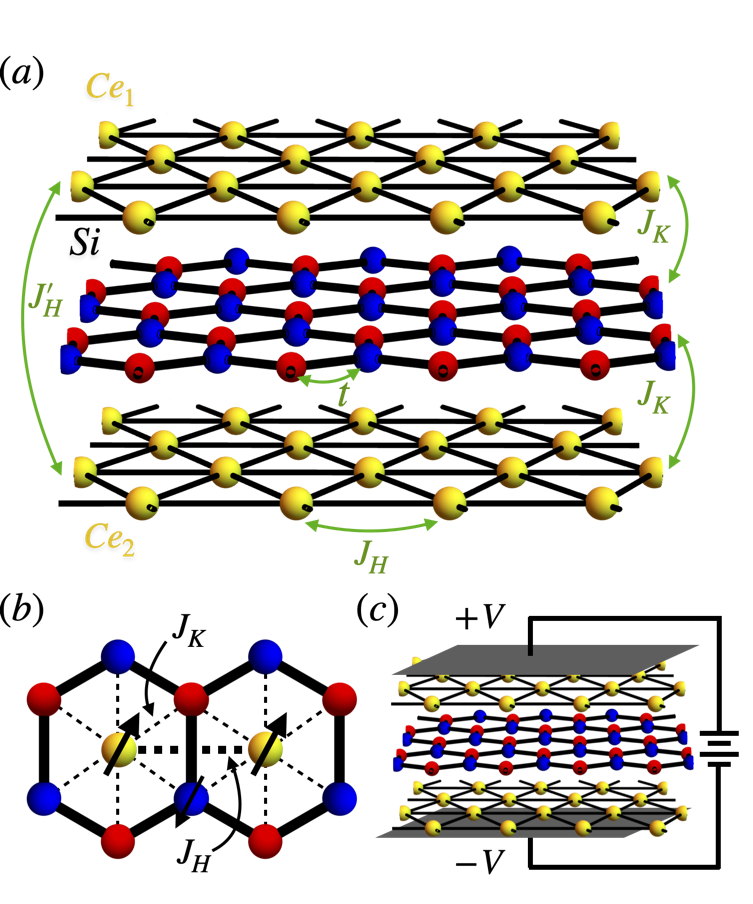}
    \caption{(a) Schematic of the model: Ce sites (yellow) form two layers of triangular lattice. The silicene layer -- a honeycomb lattice formed by Si atoms (red and blue for the A and B sublattice)-- is sandwiched between the Ce layers. The arrows indicate the interaction terms in our model including the Kondo interaction ($J_K$) between the local moments and the conduction electrons as well as intra ($J_H$) and interlayer ($J_H^\prime$) Heisenberg exchange between the local moments. Iodine layers do not play a significant role in the low energy model and therefore they are omitted on the schematic. (b) Top view of the model that shows the local moment interacting with 6 conduction electron sites via Kondo interaction. %Unlike the standard RKKY interaction where the local moments couple to different conduction electron sites, in our model the NN local moments couple to the same Si site, which results in a predominant FM NN RKKY interaction. 
    (c) Electric field tuning of the monolayer: applying +V, -V potential, shifts the energy levels of the Ce layers, increasing the Kondo coupling on one layer while decreasing on the other. Electric field does not affect the chemical potential of the conduction electrons.}
    \label{Fig:1}
\end{figure}
Heavy fermions are an archetypical class of strongly correlated materials that possess local moments, generally originating from lanthanide or actinide ions, interacting with conduction electrons via an antiferromagnetic Kondo interaction\cite{Stewart_RMP1984, Wirth_NatRevMat2016, Coleman_book}. They host an array of exotic phenomena including unconventional superconductivity, hidden order and strange metals\cite{White_PhysicaC2015, Mydosh_JPCM2020, Coleman_JPCM2001}. Most heavy fermion materials are intermetallics and exhibit strong chemical bonding along all directions which prevent the isolation of two-dimensional (2D) monolayers. In contrast, the recent discovery of magnetism and 2D heavy quasiparticles in CeSiI\cite{Okuma_PRM2021, Posey_Nature2024} provides a rare example of a van der Waals (vdW) heavy fermion platform. 2D vdW materials are highly tunable systems that can be controlled via strain, gating and electric field\cite{Liu_NatRevMat2016}. Furthermore, they can be arranged in different stacking patterns and twisted to form moir\'e superlattices which can give rise to even richer phenomena that may not be possible to realize in monolayers\cite{He_ACSNano2021}. Therefore, introducing heavy fermions to 2D materials adds a new dimension to the list of phenomena, functionalities and the potential for vertical integration in vdW heterostructures.

CeSiI orders magnetically at T$_c$ = 7.5 K. Neutron scattering experiments\cite{Okuma_PRM2021} show that the ordering wave vector is ${\bf q} = (0.28, 0, 0.19)$, resulting in a spiral ground state with a long wavelength, $\lambda = 2\pi/q_a \sim 22$ unit cells, considering a monolayer. Furthermore, the Sommerfeld coefficient of CeSiI, $\gamma_{\rm CeSiI}$= 0.125 Jmol$^{-1}$K$^{-2}$, shows significant enhancement compared to its non-magnetic analog LaSiI, $\gamma_{\rm LaSiI}$= 0.003 Jmol$^{-1}$K$^{-2}$, indicating formation of heavy fermions\cite{Posey_Nature2024}. The temperature dependence of resistivity, Fano shape of the tunnelling spectrum and the angle resolved photoemission spectroscopy  experiments further corroborate the existence of heavy quasiparticles\cite{Posey_Nature2024}.

\begin{figure}[t]
    \centering
    \includegraphics[width=1\linewidth]{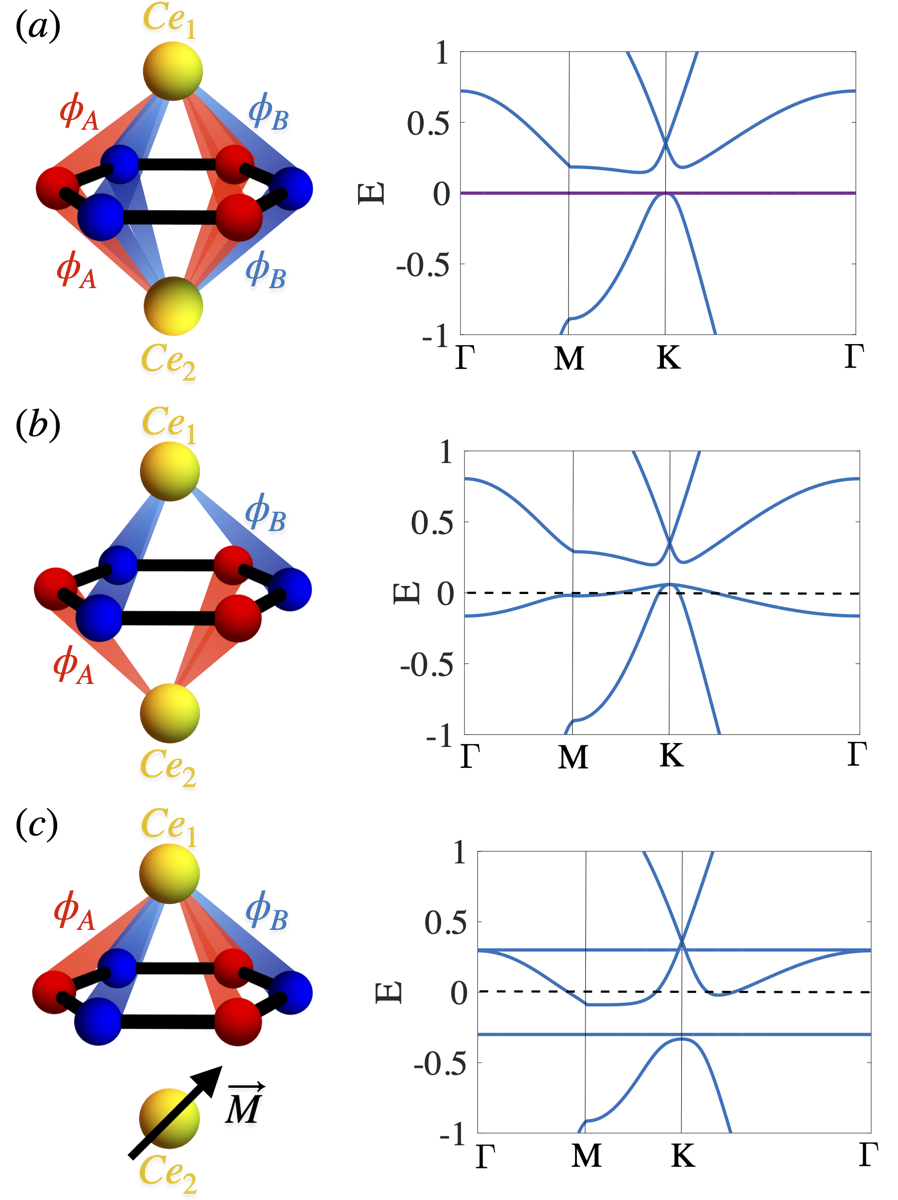}
        \caption{Schematic of different phases and the corresponding excitation spectra: (a) a uniform HFL phase where all $\phi$'s are equal. This phase has flat bands at zero energy and therefore it is unstable. (b) A nematic heavy Fermi liquid where local moments in layer 1 primarily couple to one sublattice whereas the layer 2 couples to the other sublattice. (c) A coexistence of magnetic order and a uniform heavy Fermi liquid.}
        \label{Fig 2}
\end{figure}

Motivated with these advancements, we develop an effective model for a monolayer CeSiI that involves the conduction electrons at the silicene layer -- a honeycomb lattice formed by Si atoms -- and the local moments at the Ce sites. Ce ions form two layers of triangular lattices that sandwich the silicene layer. Each Ce site sits at the center of the honeycomb lattice and couples to six NN conduction electron sites as depicted in Fig.~\ref{Fig:1}. We solve this model with the extended Kondo interactions using mean field theory of Abrikosov fermions. Our main results are as follows: (i) since both Ce layers couple to the conduction electrons via the same form factor, a uniform heavy fermion state only screens the symmetric superposition of the local moments on each layer while the antisymmetric  superposition remains unscreened. This results in an unstable phase with flat bands at the chemical potential. (ii) Via a nonrestrictive mean field ansatz, we find that the channel symmetry can be spontaneously broken and two types of heavy fermion phases can be stabilized. The first one is a non-magnetic phase where local moments on layer 1(2) couple primarily to the silicene sublattice A(B) as depicted in Fig.~\ref{Fig 2}(b). This phase preserves inversion symmetry but breaks mirror reflection and C$_6$ rotational symmetry and thus exhibits a nematic order parameter. The second phase breaks the channel symmetry by ordering magnetically on one layer and forming a uniform heavy fermion on the other layer resulting in a coexistence similar to experimental observations. For large Heisenberg interaction, both layers order magnetically. (iii) We show that the stability of the coexistent phase can be enhanced further via the application of an external electric field.

The rest of the paper is organized as follows. In Sec. II, we introduce the model, discuss the origin of the ferromagnetic (FM) nearest neighbor (NN) exchange and describe the methodology including the mean field theory. In Sec. III, we present our results for different parameter regimes. In Sec. IV, we conclude with a summary of our results and an outlook. 

\section{Model and Methods}
The first principles calculations suggest that the conduction electrons primarily occupy the silicene bands\cite{Jang_npj2DMat2022, Fumega_arXiv2024}. These bands are self-doped by the Ce d-electrons, shifting the chemical potential far from the Dirac point. We consider a model where the Ce local moments on layer 1, 2 couple to the conduction electrons at six silicene sites via Kondo interaction.  In addition, we consider a FM intra- and interlayer Heisenberg interactions among the local moments. The Hamiltonian reads,
\begin{eqnarray}
H&=&-t\sum_{\langle ij \rangle \sigma} (c_{iA\sigma}^\dagger c_{j B\sigma}+ {\rm H. c.})-\mu\sum_{i\sigma s}c^\dag_{is\sigma}c_{is\sigma}\nonumber \\&&+ \sum_{i,\delta, \nu} J_{K}^\nu\boldsymbol\sigma_{fi\nu} \cdot \boldsymbol\sigma_{c (i+\delta)} -J_H\sum_{\langle ij \rangle \nu} \boldsymbol\sigma_{fi\nu} \cdot \boldsymbol\sigma_{fj\nu}\nonumber \\
&& - J_H^\prime \sum_i \boldsymbol\sigma_{fi1} \cdot \boldsymbol\sigma_{fi2} 
\label{eq:H}
\end{eqnarray}
where $\nu=1,2$ is the Ce layer index, $s=A,B$ represent the silicene sublattice and $\delta$ denotes the six nearest Ce-Si neighbors. $c^\dagger$ and $\boldsymbol\sigma_{c(f)}$ are the creation and spin operators of the conduction electrons (local moments). $J_K^{1(2)}$ is the Kondo coupling in layer 1(2), $J_{H}$ and $J_{H}^\prime$ are the intra and interlayer Heisenberg exchange respectively. Eq.~\ref{eq:H} can be derived from a Schrieffer-Wolff transformation of the periodic Anderson model\cite{Schrieffer_PR1966}. Note that this procedure gives rise to additional terms such as a correlated hopping combined with a spin-flip\cite{Alexandrov_PRB2014}. However, we are restricting our analysis only to the spin-spin interactions. Note that extended Kondo interaction is crucial to obtain topological phases in heavy fermion systems\cite{Dzero_ARCMP2016, Alexandrov_PRL2015, Ghazaryan_NJP2021, Lai_PNAF2018}. It has also been investigated in the context of d-wave heavy fermions\cite{Ghaemi_PRB2007} and unusual magnetic interactions\cite{Ahamed_PRB2018}.

%Firstly, we briefly discuss a possible origin of the dominant FM Heisenberg exchange among the local moments. As depicted in Fig.~\ref{Fig:1}(b), the NN local moments can couple to the same conduction electron sites. Therefore, the primary contribution to the Ruderman-Kittel-Kasuya-Yosida (RKKY) interaction\cite{Ruderman_PR1954, Yosida_PR1957, Kasuya_PTP1956} originates from the processes where NN local moments polarize the same silicene site, which in return give rise to a FM Heisenberg exchange.

The spiral ground state of CeSiI with a long wavelength, $\lambda \sim 22a$ indicates that the the angle between the magnetic moments within a unit cell is quite small ($\Delta \theta \sim 16^{\circ}$) and locally, they can approximately be considered FM. One of the simplest ways to generate such a spiral is to consider a spin Hamiltonian with a large FM NN exchange ($J_1$) and a weak antiferromagnetic AFM next nearest neighbor exchange ($J_2$). Performing a Luttinger-Tisza analysis\cite{Luttinger_PR1946} on the $J_1-J_2$ Hamiltonian results in $H= \sum_{q}J({\bf q})S_{{\bf q}}S_{{\bf -q}}$ where $J({\bf q})=-J_1\sum_\Delta \cos{\mathbf{q}\cdot\mathbf{\Delta}}+J_2\sum_{\Delta'} \cos{\mathbf{q}\cdot\mathbf{\Delta}'}$, where we sum over $\Delta$ and $\Delta'$ which are the NN and NNN vectors respectively. For $J_2/J_1 \simeq -0.11$, it is possible to obtain a spiral ground state with {\bf q} that is similar to the experimental value. For the remainder of the manuscript, we only consider a FM NN Heisenberg exchange to avoid the intricacies arising from a large unit cell.

%In heavy fermions, the magnetic interaction among the local moments are mainly mediated via the Ruderman-Kittel-Kasuya-Yosida (RKKY) interaction\cite{Ruderman_PR1954, Yosida_PR1957, Kasuya_PTP1956}. RKKY interaction involves calculating the spin-dependent polarization diagram and therefore oscillates with the Fermi momentum and decay as a power-law at long distances, $J_{RKKY}(r\gg 1)\sim J_K^2/W\cos(k_F r)/r^D$ where $W$ and $D$ are the bandwidth and the dimension respectively. Intuitively, a local moment polarizes the conduction electrons, which propagate in space and undergo Friedel oscillations and in return interacts with a local moment at a different site. However, in our model, the nearest neighbor local moments couple to the {\it same} conduction electron. As depicted in Fig. X(x), since the nearest neighbor local moments polarize the same conduction electron, it is natural that the dominant Heisenberg interaction is FM. This argument can be justified by calculating the polarization bubble with vertices at the same site.

Next, we describe the mean field approach.  
%to solve this model and investigate the competition between magnetism and heavy Fermi liquid behaviour. 
We begin by representing the local moment spins in terms of Abrikosov fermions\cite{Coleman_JoPCM1989, Senthil_PRB2004, Pixley_PRL2014, Guerci_SciAdv2023}: $\boldsymbol\sigma_{fi\nu}=f^\dag_{i\nu\sigma}\boldsymbol\sigma_{\sigma\sigma'}f_{i\nu\sigma'}$, subject to the constraint $\sum_{\sigma}f^\dag_{i\nu\sigma}f_{i\nu\sigma}=1$. We decouple the Kondo interaction term in the hybridization channel: $(f^\dag_{i\nu\alpha}\boldsymbol\sigma_{\alpha\beta}f_{i\nu\beta})(c^\dag_{i+\delta\alpha'}\boldsymbol\sigma_{\alpha'\beta'}c_{i+\delta\beta'})= -2(f\dag_{i\nu\alpha}c_{i+\delta \alpha})(c^\dag_{i+\delta \beta}f_{i\nu\beta}) \approx \sum_{s}\phi_{\nu s}(\sum_{\delta_s}c^\dag_{i+\delta_s s \sigma})f_{i\nu \sigma} +\rm H.c.$,
where $s=A,B$ and $\delta_s$ are vectors connecting the center of the hexagon to the $s$ sublattice. The Kondo hybridization is $\phi_{\nu s}=\langle f^\dag_{i\nu \sigma}(\sum_{\delta_s}c_{i+\delta_s s \sigma})\rangle/3=\langle \Gamma_s(k)f^\dag_{k\nu\sigma}c_{ks\sigma}\rangle/3$, where $\Gamma_s(k)=\sum_{\delta_s}e^{i\boldsymbol{k}\cdot\delta_s}$. We decouple the Heisenberg interaction $\sigma^\alpha_{fi\nu}\sigma^\alpha_{fj\nu}=M^\alpha\sigma^\alpha_{fj\nu} + M^\alpha\sigma^\alpha_{fi\nu}$ where $M^\alpha_\nu=\langle \sigma^\alpha_{fi\nu} \rangle=\langle f^\dag_{i\nu}\sigma^\alpha f_{i\nu}\rangle$. 
The mean field Hamiltonian reads,
\begin{eqnarray}
H_{MF}&=&-t\sum_{k \sigma} (c_{kA\sigma}^\dagger c_{k B\sigma}+ {\rm H. c.})-\mu\sum_{k\sigma s}c^\dag_{ks\sigma}c_{ks\sigma}\nonumber \\&&-\sum_{k, \nu} 2J^\nu_K\sum_{s}\phi_{\nu s}\Gamma^*_s(k)c^\dag_{k s \sigma}f_{k\nu \sigma} +\rm H.c. \nonumber \\
&&-3J_H\sum_{ k \nu} \boldsymbol{M}_\nu\cdot (f^\dag_{k\nu}\boldsymbol\sigma f_{k\nu}) \nonumber \\
&&- J_H^\prime\sum_{ k \nu} \boldsymbol{M}_1\cdot(f^\dag_{k2}\boldsymbol\sigma f_{k2})+\boldsymbol{M}_2\cdot(f^\dag_{k1}\boldsymbol\sigma f_{k1})\nonumber \\
&&-\sum_{k\nu\sigma}\lambda_\nu(f^\dag_{k\nu\sigma}f_{k\nu\sigma}-1) 
\label{eq:Hmf}
\end{eqnarray}
We estimate the mean field order parameters $\{\phi_{1A}, ~\phi_{1B},~ \phi_{2A}, ~\phi_{2B},~ \boldsymbol{M}_1, ~\boldsymbol{M}_{2} \}$ and the Lagrange multipliers $\{ \lambda_1, \lambda_2 \}$ self consistently. For the remainder of the manuscript we fix $J^\nu_K/t=0.65$ (the bandwidth is $6t$) and the chemical potential $\mu$ sufficiently away from the Dirac points, resulting in $n_c=\sum_{i\sigma }c^\dag_{i\sigma}c_{i\sigma}=0.82$.

Apart from the ground state phase diagram, we also explore the effects of electric field tuning. We consider a setup where $+V$ and $-V$ voltages are applied by top and bottom gates. This does not affect the chemical potential of the silicene layer. However the energy levels of the local moments on layer 1(2), $\epsilon_{f1(2)}$, increase (decrease) with respect to the chemical potential due to the potential drop, $\epsilon_{f1/2} \rightarrow \epsilon_{f1/2}\mp e\Delta V$. Since Ce ions primarily fluctuate via $f^1 \leftrightarrow f^0$, the Kondo exchange on two layers gets modified as follows, $J_K^{1/2} = t^2_{cf}/(\epsilon_f\mp e\Delta V) = J_{K0}/ (1\mp e\Delta V/\epsilon_f)$ where $J_{K0}=t^2_{cf}/\epsilon_f$ is the Kondo coupling in the absence of an external electric field. 

\section{Results and Discussion}
There are two key ingredients in our model that differentiate it from a standard Kondo-Heisenberg model. The first one is the extended Kondo exchange where each local moment couple to six conduction electrons. The second is the presence of two layers of local moments that couple to the conduction electrons via the same form factor. To illustrate the impact of the latter, we consider our model with only a single layer of local moments. Our mean field analysis shows that a uniform heavy Fermi liquid (HFL) form with $\phi_A = \phi_B$ and $\boldsymbol{M}=0$ for small $J_H/J_K$. A first order transition to a magnetically ordered state takes place at $J_H/J_K = 0.52$. Similar behavior has been observed in other Kondo-Heisenberg type models\cite{Guerci_SciAdv2023}. Note that the uniform HFL has the same $\phi$ for all 6 bonds and carry zero magnetization whereas the magnetically ordered phase has $\phi=0$ for all bonds.

Proceeding with the original model with two layers of local moments, we recognize that a uniform HFL with $\phi_{1A}= \phi_{1B}= \phi_{2A} = \phi_{2B}$ is not attainable in this case. As shown in Fig.~\ref{Fig 2}(a), a uniform HFL has doubly degenerate flat f-bands at zero energy. This phase is unstable to perturbations and can not be the ground state. To elucidate the origin of the flat bands, we perform a rotation to a basis with the symmetric and antisymmetric linear combinations of the $f$ fermions as  $f_{p\sigma}^\dagger = (f_{1\sigma}^\dagger +f_{2\sigma}^\dagger)/\sqrt{2}$ and $f_{m\sigma}^\dagger = (f_{1\sigma}^\dagger -f_{2\sigma}^\dagger)/\sqrt{2}$. For uniform HFL with equal $\phi$'s for all bonds, it is clear that in eq. \ref{eq:Hmf} only the symmetric linear combination couples to the conduction electrons whereas the antisymmetric combination remains unscreened, resulting in flat bands at E=0. Therefore, the uniform HFL solution shows resemblance to an underscreened Kondo model where the conduction electrons are unable to screen all of the local moments\cite{Nozieres_JP1980}. However, we note that there are exactly two local moments and two conduction electron site per unit cell and thus our model should not be underscreened. This paradox arises due to the fact that the local moments on layer 1 and 2 couple to the conduction electrons with the same form factor (equal $\phi$'s). Accordingly, only a linear combination of them gets screened. To overcome this issue, we perform unrestricted mean-field calculations where we allow $\{\phi_{1A}, ~\phi_{1B},~ \phi_{2A}~, \phi_{2B} \}$ to be different from each other in order to break the channel symmetry between the layers. We obtain two non-uniform HFL phases: (i) a non-magnetic nematic HFL and (ii) a coexistence of a uniform HFL on one layer and magnetic order on the other layer.

{\it Nematic heavy Fermi liquid.} As depicted in Fig.~\ref{Fig 2}(b), in this phase the local moments in layer 1 couple primarily to one sublattice whereas layer 2 couples to the other sublattice, such as $\phi_{1B} = \phi_{2A} \gg \phi_{1A} =\phi_{2B}$. This phase preserves the inversion symmetry whereas it breaks sublattice, mirror reflection and C$_6$ rotational symmetry. Consequently, it exhibits a nematic order parameter. Nematic HFL was first proposed by Ref.~\citenum{Rau_PRB2014} in the context of chiral spin liquids hybridizing with metals. However, in our model it is generated self-consistently. Since nematic HFL breaks point group symmetries, it exhibits a Ginzburg-Landau type phase transition with T$_c$ set by the Kondo temperature. It is one of the rare examples of a phase transition driven by Kondo effect and in that sense it is similar to hastatic order\cite{Chandra_Nat2013} and Kondo stripe order\cite{Zhu_PRL2008}.

\begin{figure}[t]
\centering
\includegraphics[width=1\linewidth]{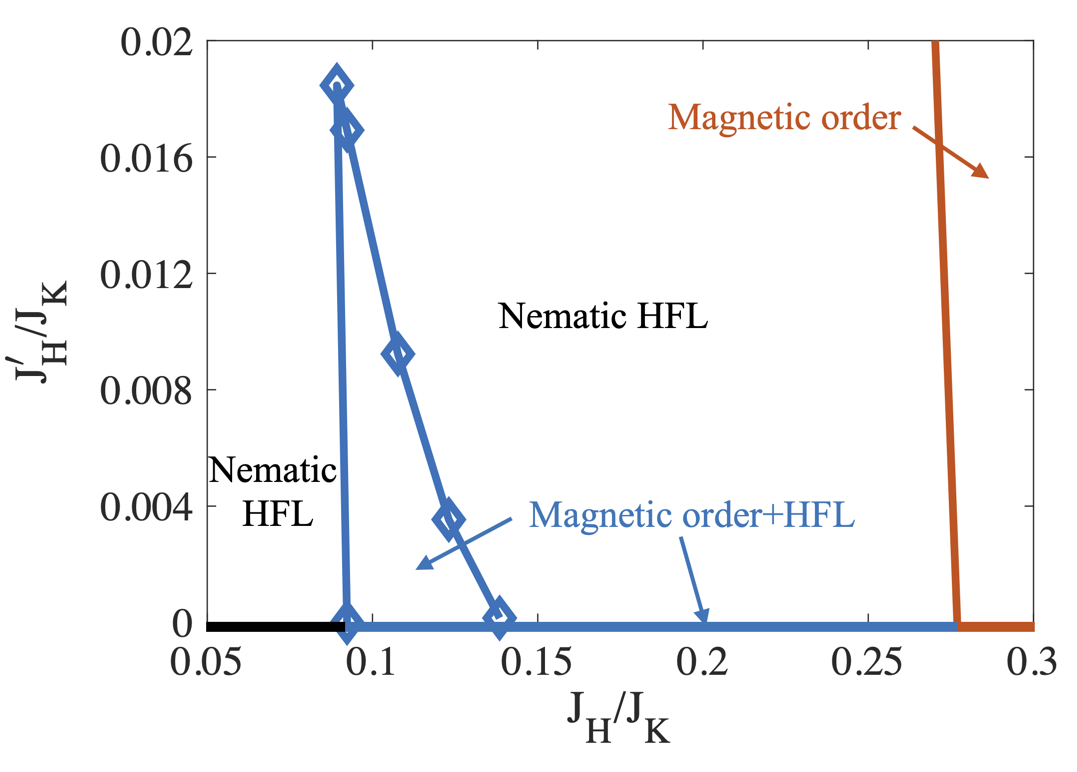}
\caption{Ground state phase diagram as a function of $J_H/J_K$ and $J_H^\prime/J_K$. We obtain a non-magnetic nematic HFL, a coexisting magnetic order and HFL and a magnetic order on both layers. All of the phase transitions are first order.}
\label{Fig 3}
\end{figure}

{\it Coexisitence of magnetic order and HFL.} This phase breaks the channel symmetry by ordering magnetically on one layer and forming a uniform HFL on the other layer such as $\phi_{1A}=\phi_{1B}\neq 0; ~\phi_{2A}=\phi_{2B}=0$ as depicted in Fig.~\ref{Fig 2}(c). Unlike spin density wave type magnetism observed in heavy fermions such as Ce$_3$Pd$_{20}$Si$_6$ \cite{Mazza_PRB2022}, the coexisting phase in our model has a clear phase separation in real space. A similar separation of magnetic order and heavy quasiparticles has been observed in modulated magnetic textures in CeRhIn$_5$\cite{Fobes_NatPhys2018}, CeAuSb$_2$\cite{Marcus_PRL2018} and also in CeSb but in momentum space\cite{Jang_SciAdv2019}.

\begin{figure}[t]
\centering
\includegraphics[width=1\linewidth]{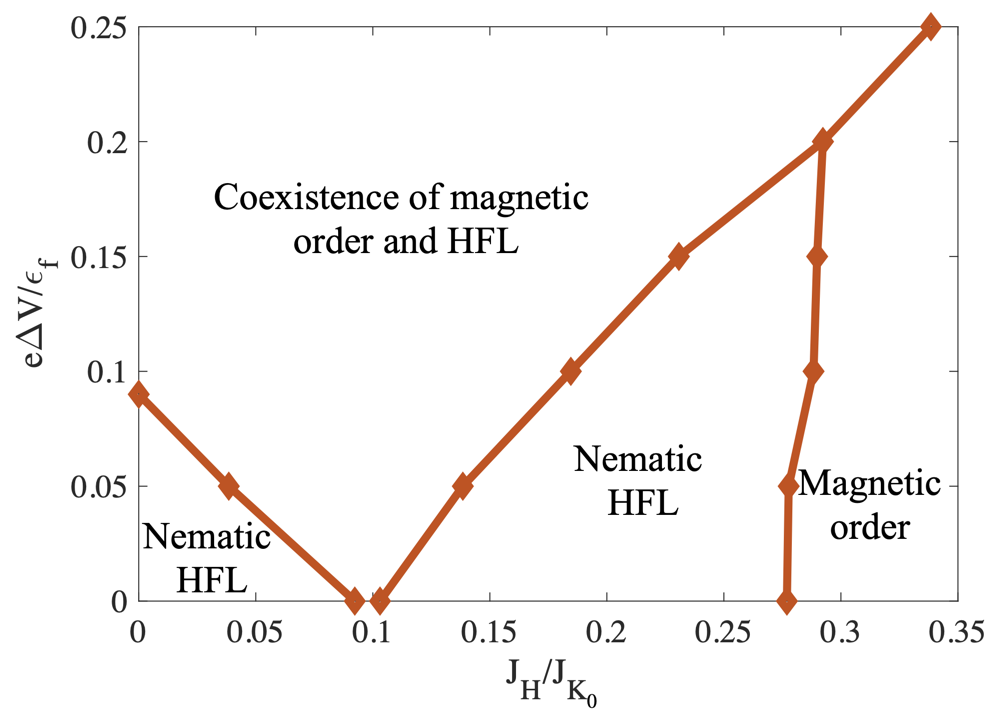}
    \caption{Effects field tuning of the phase diagram for $J_{H}^\prime/J_K = 0.01 $: Electric field shifts the energy levels of the local moments and therefore increases the Kondo interaction on one layer while decreasing it on the other layer. This significantly enhances the stability of the coexistence of magnetic order and HFL.}
    \label{Fig 4}
\end{figure}

In Fig.~\ref{Fig 3}, we present the ground state phase diagram as a function of $J_H/J_K$ and $J_H^\prime/J_K$. For $J_H^\prime=0$, the ground state evolves from a nematic HFL to a coexisting magnetic order and HFL to a polarized magnetic phase as a function of $J_H$. For small interlayer Heisenberg interaction, $J_H^\prime$, the nematic HFL is unaffected. On the contrary, for the coexistence phase, $J_H^\prime$ acts as a static magnetic field for the HFL and destabilizes it quite rapidly. All of the phase transitions are first order and are determined by the energy crossings of the corresponding phases.

Next, we discuss the electric field tuning of the phase diagram. As described in the previous section, the external electric field modifies the Kondo coupling on two layers as $J_K^{1/2}= J_{K0}/ (1\mp e\Delta V/\epsilon_f)$ while maintaining the same chemical potential for the conduction electrons. As shown in Fig.~\ref{Fig 4}, electric field enhances the stability of  the coexistence of magnetic order and HFL over the other phases. This is due to the fact that increasing the Kondo coupling on the layer where the HFL resides lowers the energy whereas decreasing the Kondo coupling on the magnetic layer does not impact the energy. In total, the energy of the coexistent phase decreases significantly. Conversely, the nematic HFL has finite hbyridization on both layers, and the modification of the Kondo couplings approximately cancel each other and the energy is not affected in significant manner. 

There are several key experimental signatures of the non-uniform heavy fermion phases that we predict. For instance, NMR experiments can distinguish if the magnetic order is uniform or if there is a phase separation of magnetism and HFL on different layers. Regarding the nematic HFL, C$_6$ symmetry breaking can be deduced from angle dependent transport, Raman spectroscopy as well as the finite temperature phase transition.

\section{Conclusions}
We constructed a low energy model for the newly-discovered vdW heavy fermion material CeSiI and studied its phase diagram via mean field theory. We showed that the unique geometry of the interactions prevents a uniform HFL. In return, we showed that a nematic HFL or a coexistence of magnetic order and HFL can be stabilized. In particular, the coexistent phase can be further enhanced by the application of an external electric field. Interesting future directions include heterostructures involving CeSiI and its moir\'e superlattices.

\section{Acknowledgements}
We thank Filip Ronning, Emilian Nica and Turan Birol for fruitful discussions. This work is supported by NSF Award No. DMR 2234352.

\bibliography{references.bib}
\end{document}